\documentclass[reprint,secnumarabic,amssymb, nobibnotes, aps, prd]{revtex4-2}
\usepackage{graphicx}
\usepackage{amsmath}
\usepackage{epstopdf}

\begin{document}

\title{Nonequilibrium distribution function in the presence of a heat flux at the interface between two crystals}%

\author{A.\,P. Meilakhs} 
\email[A.\,P. Meilakhs: ]{mejlaxs@mail.ioffe.ru}
\affiliation{Ioffe Institute, 26 Politekhnicheskaya, St. Petersburg 194021, Russian Federation }
\date{\today}%

\begin{abstract}
A one-dimensional harmonic chain model is used to study the non-equilibrium distribution function of phonons induced by a heat flux across the interface between two crystals. Conditions are derived which govern the matching of distribution functions on both sides of the interface. A generalization of the Enskog--Chapman method for calculating the Kapitza conductance is introduced. A precise relation is obtained under some simplifications. 

\medskip
This version contains improved translation, modernized notation, and corrects one mistake, that does not affect the final formulae.
\end{abstract}

\maketitle

\section{Introduction}
Recently the problem of heat transfer across the interface between two media has attracted great interest because of its significant application potential  \cite{Kid}. As it was first found by Kapitza \cite{Kap}, a heat flow from one material to another is accompanied by a temperature jump at the interface. The proportionality coefficient relating the temperature jump to the heat flux is known as the thermal boundary conductance (the Kapitza conductance). Originally temperature jump was discovered at the interface between liquid helium and solid body and this situation was investigated (see Ref. \cite{Pol} and citations there). Recent years important experimental results were obtained for thermal resistance between two crystals \cite{St, exp1, exp2, exp3}. In the present paper, we want to develop the theory for this case.

 Many authors \cite{Kh, Can, Ov, Mah}, have calculated the heat flux across the interface for given temperatures of the media assuming distribution functions of phonons be an equilibrium distribution function with a corresponding temperature. However, the existence of constant heat flux across the interface assumes the presence of the same flux in the medium, which implies that the distribution functions are out of equilibrium. In the present study, phonon distribution functions (and appropriate boundary conditions for them) are introduced in the same way as for a uniform medium in the presence of temperature gradient, i.e., by the Enskog--Chapman method \cite{Ld}. For an interface between two crystals this translates into substantial corrections to the calculated value of the Kapitza conductance.

In Ref. \cite{Jh} an analysis has been performed of vibrations in a chain consisting of two semiinfinite sub-chains with one additional link connecting them at the interface. A plane wave incident on one side of the interface produces transmitted and reflected waves. The problem of finding the amplitudes of the transmitted and reflected waves have  been solved completely. As for the heat flux, in Ref. \cite{Jh} it is derived with the following arguments. In the considered model there are two solutions at a specific frequency. They correspond to a wave incident on the interface from the left and from the right. Taking the temperature calculated with the occupation numbers of the phonons incident on the interface from the left (right) to be the temperature of the medium on the left (right), one can calculate the heat flux crossing the interface as a sum of fluxes associated with each of the phonons of the system, linearize with respect to the temperature difference, and integrate over all frequencies, to come to a heat analog of the Landauer relation. Actually, however, one cannot identify the temperatures of such states with those of the media.

\section{Incorrectness of the Landauer relation for a large transmission coefficient}

Consider the simplest model of the interface between two crystals described in Introduction. Specifically, we take a unified one-dimensional chain characterized by a set of elastic constants $\beta_L, \beta_R, \beta$, which define interaction among atoms inside the left (L) and the right (R) media and at the interface. The masses of atoms in the media are $m_L$ and $m_R$ (Fig.1). Let us neglect the vibration anharmonicity. If a plane wave of frequency $\omega$ and unit amplitude is incident on the interface, it generates a reflected wave with amplitude $A$ and a transmitted wave with amplitude $B$. We refer to such vibrations as “combined”. The properties of the solution are:
$|A_L| = |A_R| = |A|$, $|B_R|^2 = ({\rho_L v_L}/{\rho_R v_R})(1-A^2)$, $|B_L|^2 = ({\rho_R v_R}/{\rho_L v_L})(1-|A|^2)$, where indices $L$ are for waves in the left chain and $R$ for waves in the right chain; $v_L$ and $v_R$ are group velocities. These relations describe continuity of the heat flux for each mode separately. It is important that $A^2 +B_{L,R}^2 \neq 1$, i.e. squared amplitudes of reflected and transmitted waves cannot be interpreted as probabilities that waves are reflected or transmitted.

In Ref. \cite{Jh}, the heat flux is found from the Landauer relation:
\begin{equation}
q = \Delta T \frac{1}{2\pi} \int \limits_0^{\infty} {t \frac{\partial n_0}{\partial T}} \hbar \omega \ {d\omega},
\end{equation}
where $t = 1-|A|^2$ is the transmission coefficient, $T \approx T_L \approx T_R$ is the mean temperature of the medium, and $\Delta T = T_L - T_R \ll T$ is the temperature difference. Note that although the integration is performed from zero to infinity, the actual transmission coefficient vanishes for frequencies in excess of the maximum frequency of vibrations in one of the media, $\omega > \mathrm{min}[\omega_{mL}, \omega_{mR}]$. Energy transfer by phonons with
frequencies $\mathrm{max}[\omega_{mL}, \omega_{mR}]> \omega >
\mathrm{min}[\omega_{mL}, \omega_{mR}]$ is possible only with the
allowance for inelastic processes \cite{Me}.

One can readily verify that for $t \approx 1$, Eq. (1) is incorrect. Indeed, Eq. (1) is formally applicable to a uniform chain, i.e., to the case of $\beta_L = \beta_R = \beta$, $m_L = m_R$, and $a_L = a_R$. Obviously enough, in this case, $t = 1$, and the temperature will not jump at the “interface”, because there can be no temperature jumps in a uniform medium. Eq. (1) predicts, however, a finite heat flux at the temperature jump which was mentioned for the first time as far back as in Ref. \cite{Can}. It may be expected that the integrand in the “correct” relation will contain a factor of the type ${1}/{(1-t)^\alpha}$, with which a finite temperature jump in a uniform medium will induce an infinite heat flux (or, simply stated, a temperature jump in a uniform medium is impossible).

For a simpler model of two connected harmonic strings such relation was derived in \cite{Maas}, $\alpha$ occurred to be equal to $1$. A derivation of such relation based on Green’s function method was also provided and the result was exactly the same. Here such derivation is based on modified the Enskog—Chapman method \cite{Ld}. On modern application of Green’s function method for calculating Kapitza conductance see \cite{Chen}. 

\begin{figure}
\includegraphics[width=0.48\textwidth]{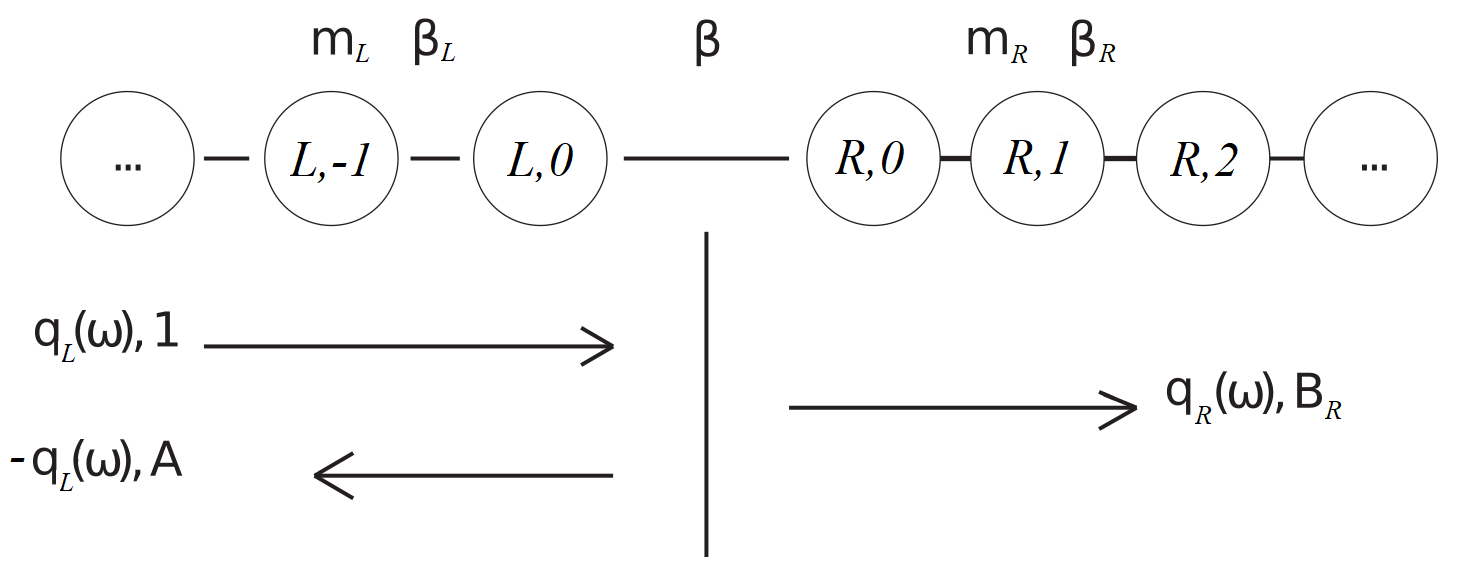}
\caption{Propagation of vibrations over two coupled semiinfinite one-dimensional chains. Indices $L$ and $R$ labels semi-infinite media, the left one and the right one.  $\beta_{L,R}$ and $m_{L,R}$ are chain coupling parameters and atomic masses in the left and the right media, respectively. $\beta$ is the chain coupling parameter at the interface which is specified by the vertical line. The left medium contains incident and reflected waves. The incident wave is assumed to have unit amplitude.  $\omega$ is the wave frequency, $q_{L,R}$ are wave vectors, and $A, B_R$ are amplitudes of reflected and transmitted waves, respectively. }
\label{fig.1}
\end{figure}

To locate the error made in the derivation of Eq. (1) let us consider a uniform chain. For a fixed frequency, it has also two solutions, a wave propagating from the left to the right and the other one, propagating from the right to the left. The state of the system at this frequency is determined by two occupation numbers $n_{\rightarrow}$ and $n_{\leftarrow}$. Following the scheme presented in the Introduction, one should define the temperatures with the relations
\begin{equation}
n_{\rightarrow} = \frac{1}{\exp{\frac{\hbar \omega}{T_1}}-1}, \ n_{\leftarrow} = \frac{1} {\exp{\frac{\hbar \omega}{T_R}}-1}.
\end{equation} 
But then, in the presence of the heat flux, we obtain $n_{\rightarrow} \neq n_{\leftarrow}$, and therefore $T_L \neq T_R$, meaning that there is a temperature jump in a uniform medium – the conclusion which, as already pointed out, is certainly unphysical. Following Enskog and Chapman, in this case, the temperature should be defined by the relation
\begin{equation}
n_0 = \frac{n_{\rightarrow} + n_{\leftarrow}}{2} = \frac{1}{\exp{\frac{\hbar \omega}{T}}-1},
\end{equation} 
and the heat flux by the relation
\begin{equation}
q = \hbar \omega v (\chi_{\rightarrow} - \chi_{\leftarrow}),
\end{equation} 
where $\chi_{\rightarrow} = n_{\rightarrow} - n_0$ and $\chi_{\leftarrow} = n_{\leftarrow} - n_0$ contain a nonequilibrium term added to the distribution function, which satisfy the condition $\chi_{\rightarrow} +\chi_{\leftarrow}=0$. In this case, the temperature along the entire chain is constant, even in the presence of the heat flux, as it should be neglecting vibration anharmonicity. Indeed, in this case there is nothing for phonons to scatter on, and the heat conductivity is infinite.

\section{Matching of the phonon distribution functions at the interface}
Let us return to the consideration of the chain with an interface and discuss again vibrations at a specified frequency. As already pointed out earlier, in this case there exist two solutions in the form of combined modes. However one cannot derive the temperature of the media from the occupation numbers of these modes; indeed, this would be equivalent to calculating the temperature from the occupation numbers of phonons propagating toward the interface only. 

Following Ref. \cite{Ov}, we could introduce a specific temperature for vibrations occurring close to the surface (i.e., for combined modes). In this approach, the temperature at the interface would differ from the temperatures in both media, and the heat would be transmitted through inelastic collisions of combined modes with conventional phonons. This method is, however, far from being physically transparent and is extremely unwieldy. An attempt to analyze the equations within this approach requires rough approximations. 

Instead, we will apply the definition of temperature by Enscog and Chapman, i.e., calculate the temperature through the arithmetical mean of the occupation numbers of phonons incident on the interface and departing from it, both from the left and from the right. To do this, we transfer from the basis of combined modes to that of conventional plane waves. Let $N_L$ and $N_R$ refer to occupation numbers of combined modes. Let $n_{L \rightarrow}$ and $n_{L \leftarrow}$ refer for occupation numbers of phonons incident on the interface and propagating from it in the left medium, while $n_{R \rightarrow}$ and $n_{R \leftarrow}$ refer to the same for phonons in the right medium.

We have to convert $N_L$ and $N_R$ to $n_{L \rightarrow}$, $n_{L \leftarrow}$, $n_{R \rightarrow}$, and $n_{R \leftarrow}$. To do this, we express the amplitudes of atomic vibrations on the left and right from the interface in terms of the creation and annihilation operators in two different ways. Let $\hat C_L$ and $\hat C_R$ denote the operators of annihilation of the combined modes, while $\hat c_{L \rightarrow}$ and $\hat c_{L \leftarrow}$ refer to phonons propagating to the right and left, accordingly, in the medium on the left. And let $\nu_L,\nu_R$ denote the normalization constants for combined modes, $\eta_L, \eta_R$ for conventional phonons in the medium on the left and right respectively. For the operator of a displacement of the $n$-th atom on the left of the interface we have:
\begin{align}
\hat u_n = \nu_L \hat C_L (e^{-i q_L a_L n} + A e^{i q_L a_L n}) +
\nu_R \hat C_R B_L e^{i q_L a_L n} = \nonumber \\
=\nu_L \hat C_L e^{-i q_L a_L n} +
(\nu_L A \hat C_L + \nu_R B_L \hat C_R) e^{i q_L a_L n},
\end{align}
and on the other hand
\begin{equation}
\hat u_n = \eta_L ( \hat c_{L \rightarrow} e^{-i q_L a_L n} + \hat c_{L \leftarrow} e^{i q_L a_L n}).
\end{equation}
Comparing Eqs. (5) and (6), we come to 
\begin{equation}
\hat c_{L \rightarrow} = (\nu_L/\eta_L)\hat C_L, \hat c_{L \leftarrow} = (\nu_L/\eta_L) A \hat C_L + (\nu_R/\eta_L) B_L \hat C_R.
\end{equation}
Now
\begin{align}
n_{L \leftarrow} = \nonumber \\ \langle n_{L \leftarrow}\ n_{L \rightarrow}\ n_{R \leftarrow}\ n_{R \rightarrow}| c^{+}_{L \leftarrow} c_{L \leftarrow}|n_{L \leftarrow}\ n_{L \rightarrow}\ n_{R \leftarrow}\ n_{R
\rightarrow} \rangle = \nonumber \\
= (\nu_L/\eta_L)^2 \langle N_L\ N_R | C^{+}_L C_L |N_L\ N_R \rangle = (\nu_L/\eta_L)^2 N_L.
\end{align}
Doing analogous calculations for all other ocupation numbers of phonons we have
\begin{equation}
\left\{\begin{aligned}
& n_{L \leftarrow} = (\nu_L/\eta_L)^2 N_L \\
& n_{L \rightarrow} = (\nu_L/\eta_L)^2 |A|^2 N_L + (\nu_R/\eta_L)^2 |B_L|^2 N_R \\
& n_{R \leftarrow} = (\nu_R/\eta_R)^2 N_R \\
& n_{R \rightarrow} = (\nu_R/\eta_R)^2 |A|^2 N_R + (\nu_L/\eta_R)^2 |B_R|^2 N_L~.
\end{aligned}\right.
\end{equation} 
We finally come to
\begin{equation}
\left\{\begin{aligned}
& n_{L \leftarrow} = |A|^2 n_{L \rightarrow} + (\eta_R / \eta_L)^2 |B_{L}|^2 n_{R \leftarrow} \\
& n_{R \rightarrow} = (\eta_L / \eta_R)^2 |B_{R}|^2 n_{L \rightarrow} + |A|^2 n_{R
\leftarrow}~.
\end{aligned}\right.
\end{equation}
Thus we obtain the matching conditions for distribution functions at the interface. Proper conditions for normalization constants $\eta_L , \eta_R$ will be introduced in the next section. We can now introduce
\begin{equation}
n_{L,R\, 0 } = \frac{n_{L,R \rightarrow} + n_{L,R \leftarrow}}{2},
\end{equation}
and on the other side
\begin{equation}
n_{L,R\, 0} = \frac{1}{\exp{\frac{\hbar \omega}{T_{L,R}}}-1}.
\end{equation}
This way we define the temperatures for the chain with the interface.

Considered from a mathematical viewpoint, the meaning of (10) is the following. The space of states of a chain vibrating at a fixed frequency is two-dimensional. Turning now to the basis of four main states, we see that those of them, which can actually be realized, form a plane in four-dimensional space defined by equations (10).

 Treated from physical point of view, (10) describe a simple situation that phonons propagating away from the interface are the sums of those that are incident on the interface from the same side and reflected and those that are incident on the interface from the other side and transmitted. The coefficients contain the squared amplitudes because the number of phonons for a given mode is proportional to the energy of vibrations, and the energy is proportional to the squared amplitude.

It is thus clear that the expansion of the basis, which could originally look rather artificial (9), has a fairly simple and informative physical meaning. It also provides a possibility to extend the definition of temperature accepted in the physical kinetics \cite{Ld} to the lattice vibrations at the interface of two crystals.

\section{Kapitza conductance}
Equations (10, 11, 12) constitute a closed system. They are solved in the reverse order. Knowing the temperatures, we find $n_{L,R\, 0}$, substitute (12) into (11) to find $n_{L \rightarrow}$ and $n_{R \leftarrow}$, then substitute the results into Eq. (10) and finally obtain $n_{L \leftarrow}$ and $n_{R \rightarrow}$. Knowing all occupation numbers, we readily find the heat flux.

Introducing the notation $\chi_{L,R} = n_{L,R \rightarrow} - n_{L,R \leftarrow}$, we come to
\begin{equation}
\begin{aligned}
\chi_L = \frac{1}{|A|^2} [(1-|A|^2)n_{L0} - (\eta_R / \eta_L)^2 |B_L|^2 n_{R0}]~, \\
\chi_R = \frac{1}{|A|^2} [(\eta_L / \eta_R)^2 |B_R|^2 n_{L0} - (1-|A|^2) n_{R0}]~.
\end{aligned}
\end{equation}
Heat flux should be equal on different sides of the interface which implies
\begin{equation}
v_L \chi_L = v_R \chi_R
\end{equation}
or
\begin{align}
v_L [(1-|A|^2)n_{L0} - (\eta_R / \eta_L)^2 |B_L|^2 n_{R0}] = \nonumber \\ = v_R [(\eta_L / \eta_R)^2 |B_R|^2 n_{L0} - (1-|A|^2) n_{R0}]~.
\end{align}
Equation (15) holds for all values of $n_{L0}, n_{R0}$, which yields
\begin{equation}
(\eta_L / \eta_R)^2 |B_R|^2 = (v_L / v_R) t~, (\eta_R / \eta_L)^2 |B_L|^2 = (v_R / v_L) t
\end{equation}
Recasting now $|B_R|^2 = t{v_L}/{v_R}$ and $|B_L|^2 = t{v_R}/{v_L}$, yields
\begin{equation}
\left(\frac{\eta_L}{ \eta_R}\right)^2 = \frac{\rho_R}{\rho_L} ~.
\end{equation}
which is the proper condition for the normalization constants. 

Substituting Eq. (16) into Eq. (13), we obtain
\begin{equation}
\begin{aligned}
\chi_L = \frac{1}{v_L} \frac{t}{1-t} [v_L n_{L0} - v_R n_{R0}]~, \\
\chi_R = \frac{1}{v_R} \frac{t}{1-t} [v_L n_{L0} - v_R n_{R0}]~.
\end{aligned}
\end{equation}
One can now conveniently transfer to a continuous limit by replacing $n_{L,R\, 0} \rightarrow n_{L,R\, 0}~{dk}/{2\pi}$, and recalling that $v={d\omega}/{dk}$, and $({d\omega}/{dk})\,n_{L,R\, 0}dk = n_{L,R\,  0}d\omega$. This allows us to linearize the expression with respect to the temperature difference, assuming the latter to be small.

To obtain the total heat flux, one has to substitute the result in Eq. (4) and integrate over frequencies. Finally we come to
\begin{equation}
q = \Delta T \frac{1}{2\pi} \int \limits_0^{\infty} {\frac{t}{1-t} \, \frac{\partial n_0}{\partial T}}\hbar \omega \ {d\omega}.
\end{equation}

We see that the prediction made at the end of the preceding section has worked out, with $\alpha=1$. We also see that for small $t$, the relation obtained translates into the Landauer relation. Equation (19) is derived under a simplifying assumption that (11) is satisfied separately for each frequency. Actually, we have here a weaker integral relation
\begin{equation}
\int \limits_0^{\omega_{m\, L,R}} n_{L,R\, 0} \hbar \omega d \omega = \int \limits_0^{\omega_{m\, L,R}} \frac{n_{L,R \rightarrow} + n_{L,R \leftarrow}}{2} \hbar \omega d \omega .
\end{equation}
Solving the problem under this condition yields the following expression for the heat flux
\begin{equation}
q = \Delta T \frac{1}{2\pi} \int \limits_0^{\infty} {\frac{t}{1-t} \, \frac{\partial n_0}{\partial T}} (1+f(\omega)) \hbar \omega \ {d\omega}.
\end{equation}
which differs from (19)  by the factor $(1+f(\omega))$ under the integral. Note that $f(\omega)\sim 1$. The factor ${1}/{(1-t)}$ is retained here.
Finding $f(\omega)$ requires combined solution of the Boltzmann equations for phonons on both sides of the interface, with the matched distribution functions (10,17)
\begin{equation}
\left\{\begin{aligned}
& n_{L \leftarrow} = |A|^2 n_{L \rightarrow} + (\rho_L/\rho_R) |B_{L}|^2 n_{R \leftarrow} \\
& n_{R \rightarrow} = (\rho_R/\rho_L) |B_{R}|^2 n_{L \rightarrow} + |A|^2 n_{R \leftarrow}
\end{aligned}\right.
\end{equation}
which is the generalization of the Enskog--Chapman method for the case of the interface between two crystals. Equation (19) may be regarded as a semi-quantitative estimate. Thus a more precise relation may be derived with generalized Chapman-Enscog method.  

\section{Comparison with experiment}
Now let us make a qualitative comparison with experiment. The results of experiments are compared with calculations based on the acoustic mismatch model (AMM) \cite{St} using Eq. (1) generalized to three-dimensional case (it involves integration over components of phonon wave vector parallel to the interface; the transmission coefficient $t$ is calculated in a more complicated way). Eq. (19) differs from (1) only by the factor ${1}/{(1-t)}$. Therefore, if $t$ does not depend on $\omega$, one can just multiply the result of calculation with AMM by the corresponding factor. At low frequencies the transmission coefficient can be derived from the theory of elasticity \cite{St}:
\begin{equation}
t(0) = \frac{4 Z_L Z_R }{(Z_L + Z_R)^2},
\end{equation}
where $Z_{L,R}$ is the so-called acoustic impedance; $Z_{L,R} = v_{L,R} \rho_{L,R}$. Because in fact the transmission coefficient decreases with increasing frequency (the fact taken into account in  \cite{St}, ${1}/{(1-t)}$ also decreases, so that the net result (Table 1) turns out to be an overestimation.

\begin{table}[h!]
\caption{}
\begin{tabular}{|r|c|c|
c|c|r|}
\hline \ & $1/(1-t)$ & Exp. & AMM & AMM(corr.) \\ \hline
Pb & 2.5 & 3 & 0.05 & 0.15 \\ \hline
Au & 50 & 4 & 0.4 & 20 \\ \hline
Al & 2.5 & 5 & 5 & 12 \\ \hline
Ti & 10 & 9 & 7 & 70 \\ \hline
\end{tabular}

\medskip
Kapitza  conductance at interface of various metals with diamond in units of $10^3 \ $W~cm$^{-2}$K$^{-1}$. Column ''AMM(corr.)'' presents the results of the AMM calculation multiplied by
$1/(1-t)$.
\end{table}

\section{Conclusions}
The data presented in the Table for gold, aluminum and titanium are overestimations, as should be expected. This may be assigned, first, to the crudness of the estimation made (see preceding Section), and, second, to the samples being not ideal. It has been shown  \cite{St}, that surface roughness causes a decrease of the Kapitza heat conductance; accordingly, calculations made without account for the roughness should result in an overestimation. The author hopes to generalize the method outlined in the present paper to the tree-dimensional case and conduct more accurate calculations.

While the value of the heat conductance for lead obtained taking into account the corrections turns out to be larger than the one calculated in terms of the AMM, it is still much smaller than that derived from experiment. This suggests that for interfaces between materials possessing substantially different acoustic impedances one should take into account inelastic energy transfer from phonons to electrons, for instance, with the inclusion of the mechanisms suggested in Refs. \cite{Mah} and \cite{Me}.

While quantitative description of Kapitza conductance is not yet found, qualitative understanding is improved. Solution of the Boltzmann equations for phonons on both sides of the interface, together with Eq. (22) gives in principal, the complete description of phenomena. Paradoxes mentioned in \cite{Can, Mah}, are resolved. Thus, the only problem is finding amplitudes of transmitted and reflected phonons in more realistic model. 

I am grateful to E.\,D. Eidelman, and D.\,G. Yakovlev for helpful discussions. I am also grateful to A.\,Ya. Vul for his interest in this work and to S.\,V. Kidalov and F.\,M. Shakhov for attracting my attention to the problem.

\end{document}